# Multiferroic properties and magnetic structure of Sm$_{1-x}$Y$_x$MnO$_3$

D. O'Flynn,[1] C. V. Tomy,[2] M. R. Lees,[1] A. Daoud-Aladine,[3] and G. Balakrishnan[1]

[1]*Department of Physics, University of Warwick, Coventry CV4 7AL, United Kingdom*
[2]*Department of Physics, Indian Institute of Technology Bombay, Mumbai 400 076, India*
[3]*ISIS Facility, Rutherford Appleton Laboratory, Didcot OX11 0QX, United Kingdom*

We have successfully induced multiferroic behavior in the A-type antiferromagnet SmMnO$_3$ by the substitution of Y at the Sm site. A magnetic transition develops at $\sim 24$ K for Sm$_{1-x}$Y$_x$MnO$_3$ ($x = 0.4, 0.5$) which is not present in the parent compound. This transition coincides with the onset of electric order, with an electric polarization measured along the $c$ axis. It is proposed that the effect of Y doping is to bring about a subtle distortion of the MnO$_6$ octahedra, causing a magnetic ordering of the Mn$^{3+}$ moments similar to that reported for the well studied multiferroic TbMnO$_3$. Following on from our previous study on polycrystalline samples, we present measurements of the magnetic and electric properties of single crystal Sm$_{0.6}$Y$_{0.4}$MnO$_3$ and Sm$_{0.5}$Y$_{0.5}$MnO$_3$. The data are summarized in a phase diagram for each of the principal crystallographic axes for the $x = 0.5$ compound. Powder neutron diffraction experiments on SmMnO$_3$ and Sm$_{0.6}$Y$_{0.4}$MnO$_3$ show that the Y substitution causes a change in the Mn-O-Mn bond angle towards the value found for TbMnO$_3$. The magnetic structure of Sm$_{0.6}$Y$_{0.4}$MnO$_3$ has been shown to consist of two phases: a sinusoidal ordering of the Mn$^{3+}$ moments below 50 K and a cycloidal ordering below 27 K. The cycloidal ordering occurs at the same temperature as the previously observed ferroelectric polarization, suggesting a similar multiferroic mechanism to that found in TbMnO$_3$.

PACS numbers: 77.80.-e, 75.60.Ej, 75.85.+t, 77.22.Ch, 77.22.Ej, 75.50.Ee, 75.25.-j

## I. INTRODUCTION

Research into multiferroic materials continues at an intense rate, motivated by fundamental questions of structure-function relationships in strongly correlated systems. The orthorhombic family of rare earth manganites, $R$MnO$_3$, has been investigated extensively since the discovery of multiferroic behavior in TbMnO$_3$, with DyMnO$_3$ and GdMnO$_3$ also exhibiting similar behavior.[1,2]

The prototype and most widely studied multiferroic manganite is TbMnO$_3$, where magnetoelectric effects have obvious potential applications for spintronics like the complete flip of the electric polarization by the application of a magnetic field of $\sim 5$ T in specific crystal directions.[1] In TbMnO$_3$, the electric polarization coincidentally appears with a change in the nature of the incommensurate magnetic order that starts developing below $T_{N1} = 41$ K. The onset of ferroelectricity coincides with a *second* magnetic transition at $T_{N2} = 27$ K at which point the Mn$^{3+}$ moments show cycloidal ordering. Between $T_{N2}$ and $T_{N1}$, the magnetic structure is sine-modulated, with the the Mn$^{3+}$ moments all oriented in the same direction, but their amplitude modulated from cell to cell. Below $T_{N2}$, this order transforms into a cycloidal magnetic order in which the moments now rotate in a crystallographic plane. Such cycloidal ordering breaks the spatial inversion symmetry and it enables the development of the spontaneous electric polarization[3]. The connection between the polarization and the cycloidal magnetic structure can be phenomenologically understood in the framework of the Landau theory, by invoking energy terms related to cross products of neighboring spins arising from Dzyaloshinskii-Moriya (DM) interactions (which vanish for any collinear structure).[4] At the microscopic level, DM-induced polarization may arise from a combination of electronic spin-current[5] or more conventional ion displacements.[6]

For neighboring spins $\mathbf{S}_i$ and $\mathbf{S}_j$ separated by a vector $\mathbf{r}_{ij}$, the electric polarization is given by[5]

$$\mathbf{P} \propto \mathbf{r}_{ij} \times (\mathbf{S}_i \times \mathbf{S}_j) \qquad (1)$$

The extent to which the steric effect and rare earth magnetism contribute to the cycloidal magnetic order is not yet clear. Tb, for instance, typically has a very large moment, which is also known to play an important role in the magnetoelectric coupling.[4,7]

Emergence of the multiferroic behavior in orthorhombic $R$MnO$_3$ manganites is achieved by reducing the size of the rare earth cation $R$ or equivalently, the tolerance factor of the perovskite structure.[8] The phase diagram as a function of $R$ reveals that only $R$=Dy shows multiferroic properties associated with an incommensurate cycloidal order similar to what was first observed for $R$=Tb. Larger rare earth ions in the $R$MnO$_3$ perovskite structure are not ferroelectric and possess the prototype $A$-type commensurate order of LaMnO$_3$, whereas smaller $R$ ions exhibit a different ground state with $E$-type magnetic order, which is still ferroelectric. For the latter, a different exchange striction mechanism is supposed to link the magnetism and the ferroelectric properties, since the $E$-type order is both commensurate and collinear so DM interactions vanish here.[9] It is clear from the phase diagram of the $R$MnO$_3$ compounds that there is a strong relation between the structural parameters and the observed magnetic and ferroelectric behavior.[8] The Mn-O-Mn bond angle, $\phi$, emerges as a key parameter, describing the local distortions produced by the tilting of the MnO$_6$ octahedra which, in turn, influence the interactions between the rare earth moments and the $Mn^{3+}$ moments. As the ionic radius of the rare earth atom is reduced, the Mn-O-Mn bond angle, $\phi$, decreases, which influences the magnetism by modifying the ratio and the frustration between first neighbor Mn-O-Mn superexchange and second neighbor Mn-O-O-Mn super-super exchange interactions. The increase in the magnetic frustration results in the emergence of a cycloidal magnetic ground state in $TbMnO_3$ and $DyMnO_3$. The manganites with larger rare earths ($R$ = La-Eu) are non-ferroelectric and exhibit $A$-type antiferromagnetic (AF) order.[10]

Multiferroic behavior can be induced in systems containing the larger rare earths when the $R$ site is doped with a smaller ion such as Y. Steric effects have recently been investigated by doping the $A$-type antiferromagnets $EuMnO_3$ and $GdMnO_3$ with smaller Y cations at the rare earth site. An investigation into $Eu_{1-x}Y_xMnO_3$, for example, has shown multiferroic behavior for $x = 0.2$.[11] Rich phase diagrams have been proposed for Y-doped $EuMnO_3$, in which the macroscopic magnetoelectric effects and the orientation of the electric polarization suggest the appearance of magnetic spiral phases where the magnetic moments rotate in different planes to $TbMnO_3$. However, no neutron scattering studies have been carried out to determine the magnetic structure of such materials. Studies of the low-temperature magnetic structures of the manganites have proved crucial to the understanding of their multiferroic properties, as seen in the case of $TbMnO_3$ where the understanding of cycloidal magnetic ordering has proved essential in explaining the properties of not just this material, but multiferroic manganites in general.

We have studied the effects of substitution at the $R$ site in $SmMnO_3$. Pure $SmMnO_3$ exhibits AF ordering with a Néel temperature of $\sim 59$ K.[12] The addition of Y on the Sm site brings about a slight distortion in the $MnO_6$ octahedra, and also reduces the effective moment of the $R$ site. As Y is non-magnetic, this approach allows the steric effect to be studied, whilst avoiding additional magnetic contributions to the system. In many respects the Y-doped $SmMnO_3$ is very similar to $EuMnO_3$. Both $SmMnO_3$ and $EuMnO_3$ have an orthorhombically distorted perovskite structure (space group P$bnm$), the only difference being that $SmMnO_3$ shows a transition to a $A$-type antiferromagnetic ordering at $T_N = 59$ K, slightly higher than for $EuMnO_3$.[13] Our previous studies on polycrystalline samples of the effect of doping $SmMnO_3$ with Y,[14] evidence was found for magnetoelectric coupling in polycrystalline $Sm_{1-x}Y_xMnO_3$ ($x = 0.4, 0.5$), from measurements of the temperature dependence of the magnetic susceptibility and dielectric constant.[14] In this communication we present the results of a detailed study on polycrystalline as well as single crystals samples of $Sm_{1-x}Y_xMnO_3$ with $x = 0.4$ and 0.5, showing the existence of multiferroic properties. These two compositions show very similar magnetic and electrical properties, and the discussions presented below apply to both compounds. We have also performed a neutron diffraction experiment in order to determine the magnetic structure of $Sm_{0.6}Y_{0.4}MnO_3$ over a temperature range spanning all the magnetic phases. The room temperature crystal structure of $Sm_{0.6}Y_{0.4}MnO_3$ and $SmMnO_3$ was also refined in order to determine the value of the Mn-O-Mn bond angle and all the structural parameters obtained are compared with those of multiferroic $TbMnO_3$.

## II. EXPERIMENTAL DETAILS

Polycrystalline $SmMnO_3$ and $Sm_{0.6}Y_{0.4}MnO_3$ were synthesized by solid state reaction. Stoichiometric ratios of high purity (99.9%) $Sm_2O_3$, $Y_2O_3$ and $MnO_2$ were thoroughly mixed, then heated at 1100°C for 12 hours. The reacted mixture was ground and heated at 1400°C two further times. The natural isotope of samarium is highly neutron absorbing, with an absorption cross section, $\sigma_a$, of 5922(56) barns.[15] To reduce this cross section, polycrystalline samples for the powder neutron diffraction experiments were prepared using isotope enriched $Sm_2O_3$ containing $^{154}Sm$ ($\sigma_a = 8.4$ barns). Powder x-ray diffraction (XRD) was conducted using a Panalytical X'Pert Pro multipurpose x-ray diffraction system with monochromated Cu $K_{\alpha 1}$ radiation. Rietveld refinements of the XRD patterns were carried out using Topas Academic software.[16] Neutron diffraction measurements were carried out at the GEneral Materials (GEM) diffractometer at ISIS,[17] with the powder samples in 6 mm vanadium cans. For $Sm_{0.6}Y_{0.4}MnO_3$, long scans ($\sim 700$ μA) were taken at base temperature and below each magnetic transition (1.5 K, 12 K and 35 K) and above the paramagnetic transition at 60 K. A temperature dependence of the magnetic diffraction was also taken, with short, 60 μA scans taken over a temperature range of 2.5 K to 58.5 K in 1.5 K increments. For $SmMnO_3$, a similar approach was taken, with 500 μA scans taken at 1.5 K, 45 K and 65 K, as well as a temperature dependence around the magnetic phase transition. Room temperature scans were taken for each compound in order to refine the crystal structure for comparison with other $RMnO_3$ compounds.

Single crystals of $Sm_{1-x}Y_xMnO_3$ ($x = 0.4, 0.5$) were grown by the floating zone method, with growth rates of 6-8 mm/hour in an atmosphere of 50% Ar / 50% $O_2$. Crystal quality and orientation were determined using the x-ray Laue technique. Crystals were cut into parallelepipeds of $\sim 3 \times 2 \times 2$ mm$^3$, with surface normals parallel to the $a$, $b$, and $c$ axes for magnetic properties measurements, and rectangular plates with an area of $\sim 2 \times 2$ mm$^2$ and $\sim 1$ mm thick for heat capacity and electric properties measurements. Electrical contacts were made by sputtering gold onto opposite faces of the crystal before connecting wires with silver conducting paste, except for the $Sm_{0.6}Y_{0.4}MnO_3$ polarization measurements, where silver paste was applied directly to the sample. Electric properties measurements were made using a multipurpose sample insert for a Quantum Design Physical

Properties Measurement System. Dielectric constant values were calculated from measurements of the sample capacitance at 10 kHz using an Agilent 4294A impedance analyser. The electric polarization data were determined from measurements of the pyroelectric current with a Keithley 6517A electrometer, using a technique described elsewhere.[1]

## III. RESULTS

### A. X-ray diffraction

X-ray diffraction patterns for polycrystalline $SmMnO_3$ and $Sm_{0.6}Y_{0.4}MnO_3$ are shown in Figure 1. In the previous study, it was found that $Sm_{1-x}Y_xMnO_3$ was single phase for compositions $x = 0 - 0.5$.[14] Table I shows the crystal structure information for the two compounds, obtained from Rietveld refinements of the x-ray diffraction data. The addition of Y on the Sm site modifies the lattice parameters of $Sm_{0.6}Y_{0.4}MnO_3$ compared to $SmMnO_3$, without altering the crystal space group. There is also a clear decrease in the mean Mn-O-Mn bond angle with Y doping. These bond angle values are approximate and neutron diffraction experiments were performed to more accurately ascertain the atomic positions of the oxygen. A more detailed analysis of the crystallographic and magnetic structure obtained using powder neutron diffraction are given in section III D.

### B. Magnetic properties

Figure 2 shows the magnetic susceptibility data for single crystal $Sm_{0.6}Y_{0.4}MnO_3$. Very similar susceptibility behavior was seen for both $Sm_{0.6}Y_{0.4}MnO_3$ and $Sm_{0.5}Y_{0.5}MnO_3$, and so the latter data has been omitted. The magnitudes of the susceptibility for both compounds are much lower than those reported for $SmMnO_3$.[14] In $SmMnO_3$ there is a small canting of the $Mn^{3+}$ moments, leading to a weak ferromagnetic signal. This canting appears to be suppressed in these crystals as a result of the Y doping. The magnetic properties of $Sm_{0.6}Y_{0.4}MnO_3$ and $Sm_{0.5}Y_{0.5}MnO_3$ are highly anisotropic, with large differences in the temperature dependence of the susceptibility for $T < 100$ K along the three crystallographic directions studied. At 49 K, features are seen in the susceptibility along each axis, with a large cusp present for $H \parallel b$. Another feature is seen in the susceptiblity for $H \parallel c$ at ∼ 24 K, providing evidence for a second magnetic phase transition which is not present for SmMnO$_3$.[12,13,18] We observed this feature in polycrystalline samples of the same composition, and it is much more pronounced in the single crystal samples.[14] The inverse magnetic susceptibility for each axis was fitted to the Curie-Weiss model to determine the total effective magnetic moment for the compound. The results of the fits are shown in Table II. Using experimentally determined magnetic moment values for Sm$^{3+}$ and Mn$^{3+}$ ions,[19] the expected moments for Sm$_{0.6}$Y$_{0.4}$MnO$_3$ and Sm$_{0.5}$Y$_{0.5}$MnO$_3$ were calculated as 5.00 $\mu_B$ and 4.97 $\mu_B$, respectively. The values for $p_{eff}$ obtained from the Curie-Weiss fits all lie within 4% of these theoretical values.

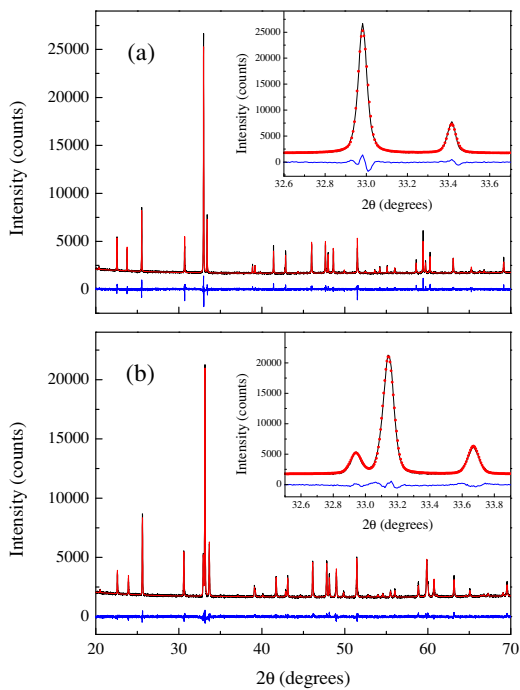

FIG. 1: (Color online). X-ray diffraction patterns for polycrystalline (a) $SmMnO_3$ and (b) $Sm_{0.6}Y_{0.4}MnO_3$ ($CuK_{\alpha 1}$ radiation); experimental data (dots), a fit to the data (line), and the difference curve. The insets show an expanded view of the high intensity peaks for both compounds.

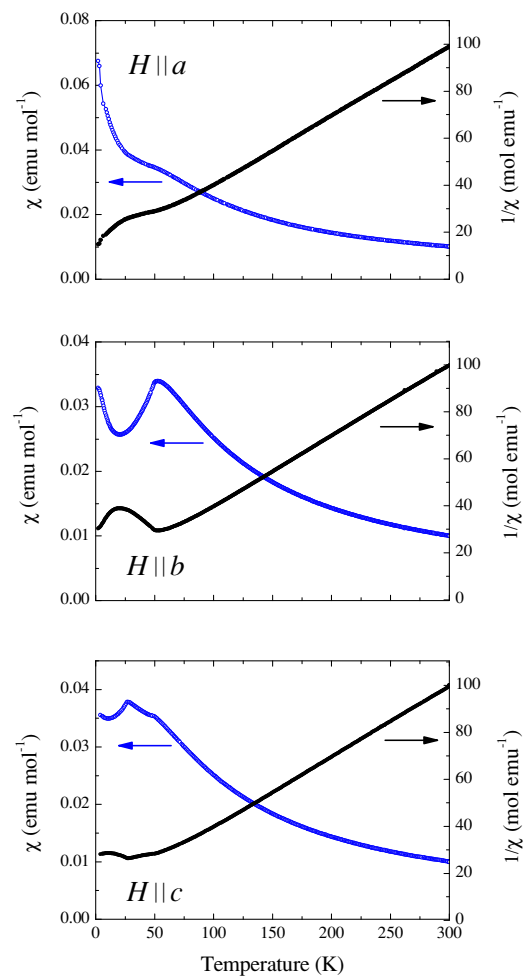

FIG. 2: (Color online). Field-cooled magnetic susceptibility (open circles) and inverse susceptibility (closed circles) for $Sm_{0.6}Y_{0.4}MnO_3$, along the $a$, $b$, and $c$ axis measured in an applied magnetic field of 5 kOe

|  | SmMnO$_3$ | Sm$_{0.6}$Y$_{0.4}$MnO$_3$ |
|---|---|---|
| $a$ (Å) | 5.36010(16) | 5.31722(5) |
| $b$ (Å) | 5.79368(14) | 5.83555(5) |
| $c$ (Å) | 7.4932(2) | 7.42946(6) |
| $V$ (Å$^3$) | 232.70(6) | 230.528(3) |
| Mn-O1-Mn (×2) | 148.4(5)° | 148.5(3)° |
| Mn-O2-Mn (×4) | 148.5(5)° | 142.1(3)° |
| ⟨Mn-O-Mn⟩ | 148.5(5)° | 144.2(3)° |
| Reliability factors | | |
| $R_p$(%) | 2.572 | 2.338 |
| $R_{wp}$(%) | 3.722 | 3.070 |
| $R_{exp}$(%) | 2.311 | 2.302 |
| $\chi^2$ | 1.610 | 1.334 |

TABLE I: Lattice parameters and Mn-O-Mn bond angles for SmMnO$_3$ and Sm$_{0.6}$Y$_{0.4}$MnO$_3$ (space group P$bnm$ for both compounds), from refined powder x-ray diffraction data.

|  | $\theta$ (K) | $p_{eff}$ | Fit range (K) |
|---|---|---|---|
| Sm$_{0.6}$Y$_{0.4}$MnO$_3$ | | | |
| $H \parallel a$ | -34.2(1) | 5.19(1) | 110-300 |
| $H \parallel b$ | -30.5(1) | 5.14(1) | 110-300 |
| $H \parallel c$ | -30.6(1) | 5.15(1) | 110-300 |
| Sm$_{0.5}$Y$_{0.5}$MnO$_3$ | | | |
| $H \parallel a$ | -32.1(2) | 5.14(1) | 120-300 |
| $H \parallel b$ | -37.2(1) | 5.16(1) | 100-300 |
| $H \parallel c$ | -33.1(1) | 5.13(1) | 110-300 |

TABLE II: Weiss temperature, $\theta$, and effective magnetic moment, $p_{eff}$, for Sm$_{1-x}$Y$_x$MnO$_3$ ($x$ = 0.4, 0.5) obtained from Curie-Weiss fits to the inverse susceptibility data.

The heat capacity data for Sm$_{0.6}$Y$_{0.4}$MnO$_3$ and Sm$_{0.5}$Y$_{0.5}$MnO$_3$ exhibit three distinct peaks, at $T_{N1}$ ∼ 47 K, $T_{N2}$ ∼ 23−25 K and $T_{N3}$ ∼ 4 K (Figure 3). The peaks at $T_{N1}$ and $T_{N2}$ are slightly lower in temperature for $x$ = 0.5 than $x$ = 0.4, and the trend is for the initial magnetic ordering temperature to decrease with increasing Y (magnetic order occurs in SmMnO$_3$ at ∼ 59 K). For each compound, the peak at $T_{N2}$ occurs at the same temperature as the corresponding feature in the magnetic susceptibility with $H \parallel c$. These results provide further evidence for a second magnetic transition at $T_{N2}$ in addition to the initial ordering at $T_{N1}$. By analogy with TbMnO$_3$ and DyMnO$_3$, the peaks at $T_{N1}$ and $T_{N2}$ are thought to be due to ordering of the $Mn^{3+}$ moments.[2] Similarly, the low-temperature ($\sim 4$) K peak at $T_{N3}$ is assumed to be due to the magnetic ordering of the $Sm^{3+}$ moments. The broad nature of this peak suggests the existence of short range ordering between the $Sm^{3+}$ ions, possibly due to the presence of Y on the Sm site interrupting the Sm-Sm exchange pathways.

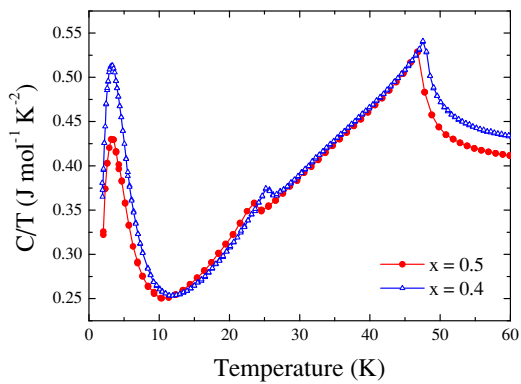

FIG. 3: (Color online). Temperature dependence of the heat capacity for Sm$_{0.6}$Y$_{0.4}$MnO$_3$ (open triangles) and Sm$_{0.5}$Y$_{0.5}$MnO$_3$ (closed circles).

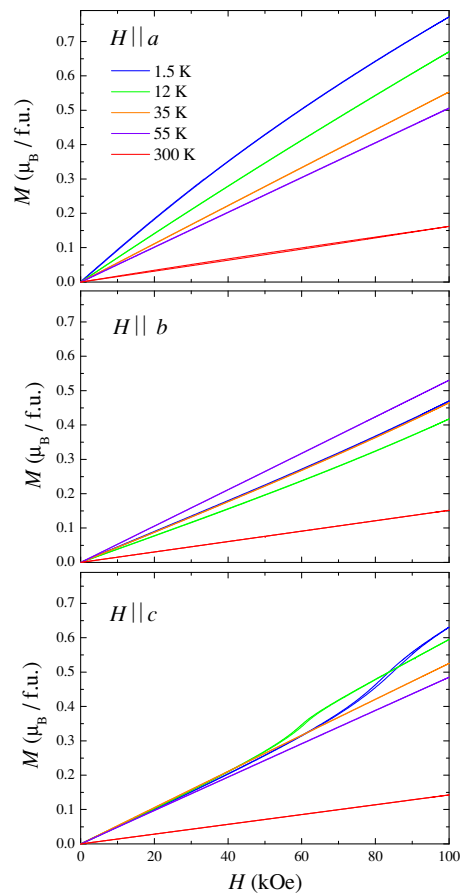

FIG. 4: (Color online). Magnetization versus magnetic field along the $a$, $b$, and $c$ axes of Sm$_{0.5}$Y$_{0.5}$MnO$_3$.

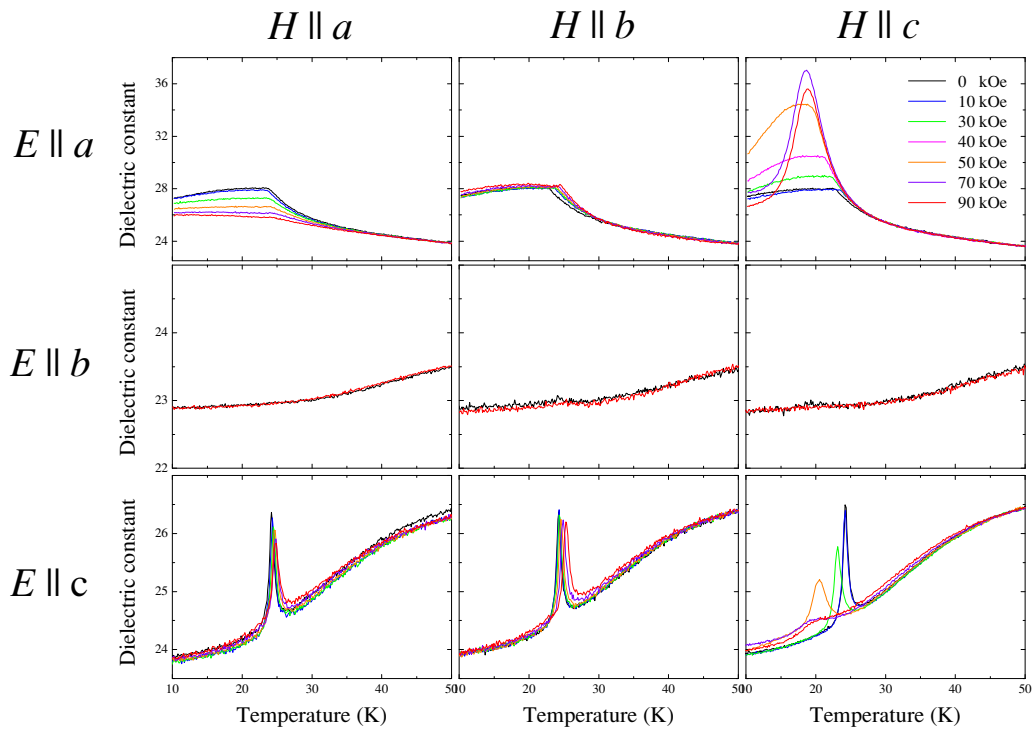

FIG. 5: (Color online). Temperature dependence of the dielectric constant along the $a$, $b$, and $c$ axes for $Sm_{0.5}Y_{0.5}MnO_3$, with magnetic fields applied along the $a$, $b$, and $c$ axes.

Figure 4 shows the magnetization along the three principal axes of $Sm_{0.5}Y_{0.5}MnO_3$ in applied magnetic fields up to 100 kOe. The theoretical saturation moments of $Sm^{3+}$ and $Mn^{3+}$ are 0.71 $\mu_B$ and 4 $\mu_B$ respectively, giving a saturation magnetization of 4.355 $\mu_B$/f.u. for $Sm_{0.5}Y_{0.5}MnO_3$. It is clear that the system is far from reaching saturation for any of the three directions of applied field. This is due to a strong antiferomagnetic coupling between the Mn moments. $a$ is the easy axis, with a downward curvature with respect the applied-field axis for 1.5 K. For $H \parallel b$ the magnetization curves upwards at higher fields. The most significant features are observed with $H \parallel c$, with metamagnetic steps present at 50 and 70 kOe at temperatures of 12 K and 1.5 K respectively (i.e. in the temperature range $T < T_{N2}$). As will be seen below, these features in the magnetization correspond to the magnetic field dependent changes in the electrical properties of $Sm_{0.5}Y_{0.5}MnO_3$.

### C. Dielectric and polarization measurements

Measurements of the dielectric properties of $Sm_{0.5}Y_{0.5}MnO_3$ show interesting features which are not present in paraelectric $SmMnO_3$. In zero magnetic field, a sharp peak is seen in the dielectric constant along the $c$ axis, coinciding with the feature in the magnetic susceptibility at $T_{N2}$. A broad feature is also seen in the dielectric constant measured along the $a$ axis below this temperature. Only a relatively small peak is seen along the $b$ axis, which is thought to be due to twinning between the $a$ and $b$ axes of the crystal. Applying a magnetic field along the $c$ axis of $Sm_{0.5}Y_{0.5}MnO_3$ results in significant changes in the dielectric constant. A field of 7 kOe causes the feature seen for $E \parallel a$ in zero field to become a large, broad peak (Figure 5(c)), and almost totally suppresses the sharp peak which is measured with $E \parallel c$ (Figure 5(i)). As with the magnetization data, applying magnetic fields along the $a$ and $b$ axes result in only small changes in the dielectric constant.

Electric polarization measurements were made along the $c$ axis below $T_{N2}$ for $Sm_{0.6}Y_{0.4}MnO_3$ (see Figure 6) and $Sm_{0.5}Y_{0.5}MnO_3$ (data not shown). As with the dielectric constant, the application of a magnetic field of 7 kOe parallel to the $c$ axis greatly decreases the magnitude of the polarization. This magnetic field also results in an increase in the polarization along the $a$ axis, indicative of a polarization flop. It should be noted that the increase in polarization along the $a$ axis is relatively

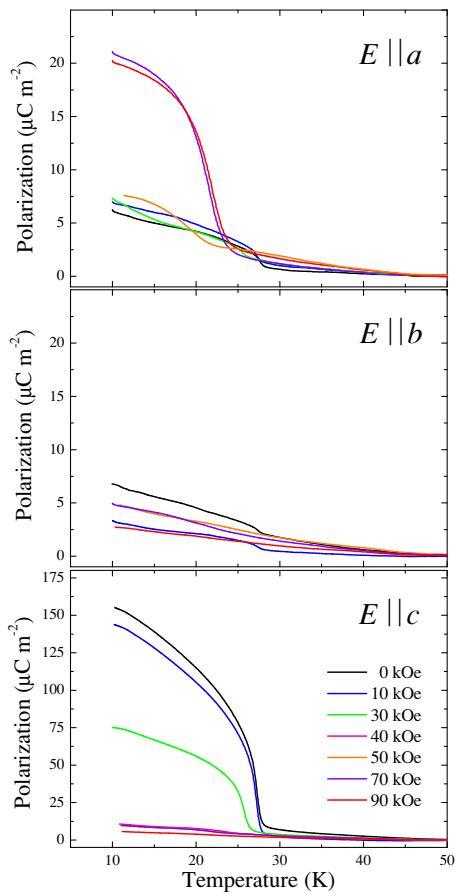

FIG. 6: (Color online). Electric polarization along the $a$, $b$, and $c$ axes of $Sm_{0.6}Y_{0.4}MnO_3$, with magnetic fields applied along the $c$ axis.

small - just over 10% of that measured along the $c$ axis in zero field. The results presented in the sections above are summarized in a phase diagram for one of the compositions, $Sm_{0.5}Y_{0.5}MnO_3$ in Figure 7.

### D. Powder neutron diffraction

#### 1. Crystallographic structure

Rietveld refinements of the diffraction patterns from GEM banks 3, 4 and 5 were carried out using the FullProf suite of programs.[20] The results of the refinement of the crystal structures of $SmMnO_3$ and $Sm_{0.6}Y_{0.4}MnO_3$ are given in Table III. Both compounds were found to have an orthorhombically distorted perovskite (space group P$bnm$) crystal structure, as found in $TbMnO_3$. It can be seen for $Sm_{0.6}Y_{0.4}MnO_3$ that there is a distortion of the unit cell due to the $Y^{3+}$ doping, with a decrease in the $a$ and $c$ lattice parameters, and an increase in the $b$ lattice parameter compared with $SmMnO_3$. The value for the mean Mn-O-Mn bond angle of $SmMnO_3$ is in good agreement with previously reported values.[21] The mean bond angle has significantly decreased in $Sm_{0.6}Y_{0.4}MnO_3$ to 145.79(1)°, bringing it close to the value for $TbMnO_3$ of 145.26(4)° (at 295 K) reported by Alonso et al.[22]

#### 2. $SmMnO_3$ magnetic structure

The magnetic structure of $SmMnO_3$ was found to be that of an $A$-type antiferromagnet below $\sim 60$ K, with the magnetic spins aligned along the $b$ axis and antiparallel arrangement along $c$ axis. The evolution of the magnetic moment can be seen in Fig. 9. This result is in accordance with previous work.[13] There were no other features observed in the magnetic properties below $T_N$.

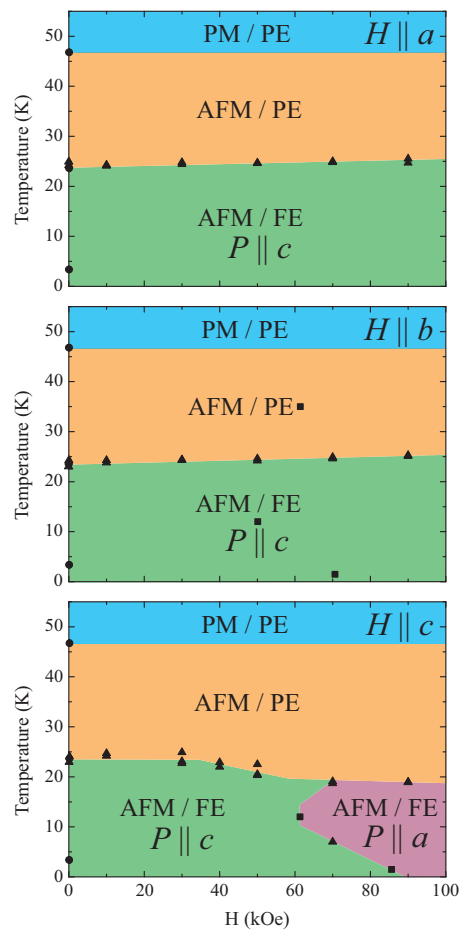

FIG. 7: (Color online). $H-T$ phase diagrams for $Sm_{0.5}Y_{0.5}MnO_3$. Regions where the samples are paramagnetic (PM), paraelectric (PE), antiferromagnetic (AFM), and ferroelectric (FE) are marked for magnetic fields applied along the $a$, $b$, and $c$ axes. Closed triangles are taken from the dielectric data, while the closed circles and closed squares are taken from the heat capacity and magnetization data respectively.
6

|  |  | SmMnO$_3$ | | Sm$_{0.6}$Y$_{0.4}$MnO$_3$ | |
|---|---|---|---|---|---|
|  |  | 1.5 K | 300 K | 1.5 K | 300 K |
| $a$ (Å) |  | 5.3584(6) | 5.3548(6) | 5.3118(5) | 5.3141(5) |
| $b$ (Å) |  | 5.7959(6) | 5.8131(7) | 5.8290(6) | 5.8432(5) |
| $c$ (Å) |  | 7.4608(8) | 7.4771(9) | 7.4047(7) | 7.4219(7) |
| $V$ (Å$^3$) |  | 231.71(4) | 232.75(5) | 229.27(4) | 230.46(4) |
| Sm/Y | 4c ($x\ y\ \frac{1}{4}$) | | | | |
| $x$ |  | 0.98588(19) | 0.98531(19) | 0.98397(16) | 0.98383(13) |
| $y$ |  | 0.07339(14) | 0.07426(13) | 0.07870(11) | 0.07918(9) |
| $B$ (Å$^2$) |  | 0.338(19) | 0.60(2) | 0.170(15) | 0.385(15) |
| Mn | 4b ($\frac{1}{2}\ 0\ 0$) | | | | |
| $B$ (Å$^2$) |  | -0.01(3) | 0.11(3) | 0.03(3) | 0.19(2) |
| O1 | 4c ($x\ y\ \frac{1}{4}$) | | | | |
| $x$ |  | 0.0949(2) | 0.0951(2) | 0.10196(19) | 0.10210(15) |
| $y$ |  | 0.4736(2) | 0.4734(2) | 0.46910(18) | 0.46846(15) |
| $B$ (Å$^2$) |  | 0.22(2) | 0.40(3) | 0.23(2) | 0.399(19) |
| O2 | 8d ($x\ y\ z$) | | | | |
| $x$ |  | 0.70725(17) | 0.70766(18) | 0.70454(15) | 0.70490(12) |
| $y$ |  | 0.32114(16) | 0.32176(16) | 0.32497(13) | 0.32540(11) |
| $z$ |  | 0.04729(12) | 0.04736(12) | 0.05004(10) | 0.05023(8) |
| $B$ (Å$^2$) |  | 0.242(16) | 0.442(19) | 0.232(14) | 0.414(13) |
| Mn-O1-Mn, $\phi_1$ (×2) |  | 148.294(3) | 148.209(3) | 145.729(2) | 145.676(2) |
| Mn-O2-Mn, $\phi_2$ (×4) |  | 147.59(5) | 147.53(5) | 145.92(4) | 145.84(3) |
| $\langle\phi\rangle$ |  | 147.82(2) | 147.76(2) | 145.86(1) | 145.79(1) |
| $R_{\rm wp}$ |  | 5.28 | 6.21 | 6.76 | 4.50 |
| Bragg $R$-factor |  | 3.14 | 4.22 | 2.20 | 2.23 |
| Magnetic $R$-factor |  | 1.06 | - | 7.10 | - |

TABLE III: Refined unit cell, atomic position, and thermal parameters for SmMnO$_3$ and Sm$_{0.6}$Y$_{0.4}$MnO$_3$ (space group P$bnm$ for both compounds), where $B$ is the isotropic temperature factor. The reliability factors given are from the refinement of the data from bank 3.

### 3. Sm$_{0.6}$Y$_{0.4}$MnO$_3$ magnetic structure

TbMnO$_3$ shows a sinusoidal magnetic order of the Mn spins below $T_{N1} = 41$ K with moments oriented along the $b$ direction, and a cycloidal order below $T_{N2} = 27$ K with moments rotating in the $bc$ plane. Sinusoidal and cycloidal modulated magnetic structures are particular cases of incommensurate magnetic order with a propagation vector **k** characterized by one or two magnetic vector components respectively. The magnetic moment $\mathbf{M}_j(\mathbf{R}_l)$ of an atom at the site $j$ in different unit cells $\mathbf{R}_l$ can therefore be written as:

$$\begin{aligned}\mathbf{M}_j(\mathbf{R}_l) = &\mathbf{A}_j \cos(2\pi\mathbf{k}\cdot\mathbf{R}_l + \varphi_j) \\ &+\mathbf{B}_j \sin(2\pi\mathbf{k}\cdot\mathbf{R}_l + \varphi_j)\end{aligned} \quad (2)$$

as described and used previously in the magnetic structure determination of multiferroic FeVO$_4$.[23] The sine-modulated incommensurate order is obtained by imposing $\mathbf{B} = 0$, and for the cycloidal order **B** is orthogonal to the common direction **A** taken by all the atomic spins in the sine-modulated phase. Here, symmetry constraints between the phases $\varphi_j$ for different atoms in the cell can be additionally imposed. The analysis of the neutron diffraction single crystal data collected on TbMnO$_3$ and reported by Kenzelmann et al.,[3] showed that the **A** and **B** components are along the $b$ and $c$ crystal axis respectively, but they follow two different irreducible representations, which relate to the appearance of multiferroic behavior as the component **B** only appears below $T_{N2}$.

We found that exactly the same spin directions and symmetry constraints enable us to fit the powder neutron diffraction collected on Sm$_{0.6}$Y$_{0.4}$MnO$_3$. It is, however, important to note that using powder diffraction data, the low-temperature cycloidal solution is in fact fully degenerate with a sine-modulated order where the spins are arranged in an **A** + **B** direction, thus, a direction tilted from the $b$ axis into the $bc$ plane. This solution can be discarded, since it also leads to unrealistically large magnetic Mn moments in some crystallographic cells.

The temperature dependence of the developing magnetic moment components in Sm$_{0.6}$Y$_{0.4}$MnO$_3$ can be seen in Figure 11(a). Above $T_{N2}$, no magnetic peaks were de-

...

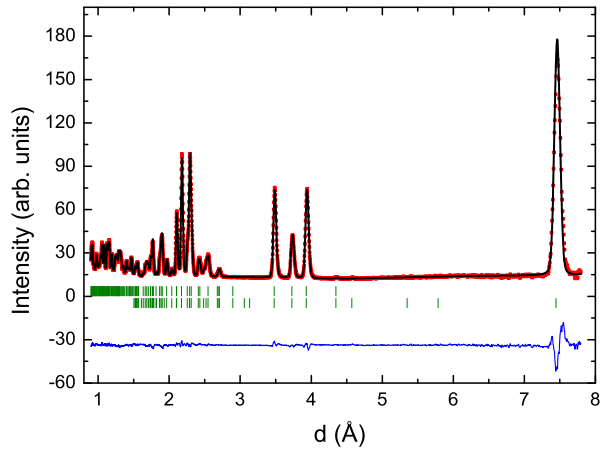

FIG. 8: Neutron diffraction pattern for SmMnO$_3$ from bank 3 of GEM, measured at 1.5 K. The red points show the data, the black line is the fit to the data, and the blue line underneath shows the difference between the two. The upper and lower green ticks indicate the positions of the nuclear and the magnetic Bragg reflections, respectively.

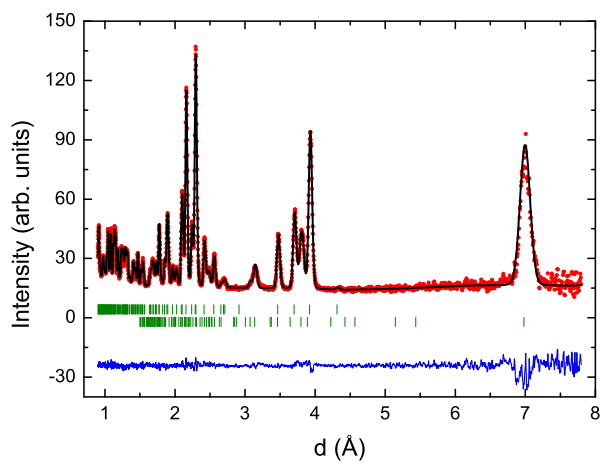

FIG. 9: Temperature dependence of the magnetic moment of SmMnO$_3$

tected, and this has been represented in Figures 9 and 11 by a magnetic moment of zero magnitude at the temperatures recorded.

As with other ortho-manganites with cycloidal magnetic order (see inset of Fig. 1 in Ref.[8]), the magnetic propagation vector $\mathbf{k} = (0\ \alpha\ 0)$ is along the $b$ direction (*Pbnm* setting). Decreasing the temperature, its varying magnitude indicates the relief of the exchange frustration through a change of the competing magnetic interactions, a process that stops when the cycloidal order is stabilized. The propagation vectors then locks below $T_{N2}$. In Sm$_{0.6}$Y$_{0.4}$MnO$_3$, $\alpha$ only slightly increases from $\sim 0.272$ at 30 K to $\sim 0.274$ at 1.5 K [Fig. 11(b)]. By comparison, the magnitude of the propagation vector of TbMnO$_3$ is 0.295 at $T_{N1}$ and decreases to $\sim 0.28$ at $T_{N2}$, after which it is nearly constant.[1] The models we have used to fit the magnetic structure of Sm$_{0.6}$Y$_{0.4}$MnO$_3$ only take into account an ordering of the Mn moments. No evidence was found from the data for the existence of Sm order in the cycloidal phase. This is in agreement with the bulk magnetization and heat capacity data presented above. We note that in TbMnO$_3$, Tb strongly influences both the magnetism and the ferroelectric properties, by showing contributions to magnetic scattering below $T_{N2}$, before eventually developing a long range order below 7 K.[24]

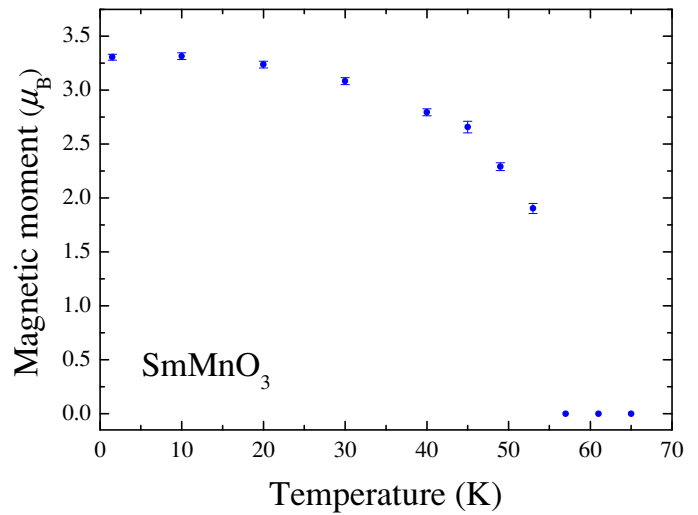

FIG. 10: Neutron diffraction pattern for Sm$_{0.6}$Y$_{0.4}$MnO$_3$ from bank 3 of GEM, measured at 1.5 K. The red points show the data, the black line is the fit to the data, and the blue line underneath shows the difference between the two. The upper and lower green ticks indicate the positions of the nuclear and the magnetic Bragg reflections, respectively.

## IV. DISCUSSION

We have identified Sm$_{0.6}$Y$_{0.4}$MnO$_3$ and Sm$_{0.5}$Y$_{0.5}$MnO$_3$ as multiferroic materials. Our structural investigations have shown that the substitution of Y on the rare earth site in SmMnO$_3$ causes a change in the crystal structure, with the Mn-O-Mn bond angle decreasing from $\sim 147.5°$ in the undoped compound to $\sim 145.8°$ in Sm$_{0.6}$Y$_{0.4}$MnO$_3$. We refer here to the bond angles obtained from the powder neutron diffraction data, as this is deemed to be a more accurate figure, given the uncertainties in fixing the lighter oxygen positions with x-ray diffraction. This bond angle value lies in the range that is found in the multiferroic $R$MnO$_3$ compounds.

The results of our magnetization, heat capacity, dielectric and polarization measurements presented for

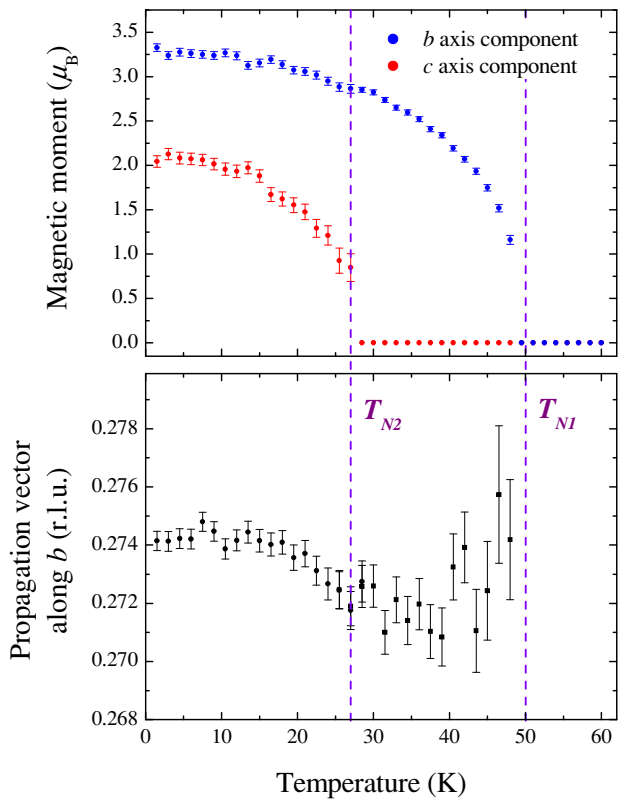

FIG. 11: (a) Temperature dependence of the $b$ axis and $c$ axis components of the Mn magnetic moment for $Sm_{0.6}Y_{0.4}MnO_3$. (b) Temperature dependence of the propagation vector along the $b$ axis.

$Sm_{0.6}Y_{0.4}MnO_3$ and $Sm_{0.5}Y_{0.5}MnO_3$ show that doping $SmMnO_3$ with Y significantly changes the magnetic properties of the system, and leads to electric properties which are not present in the parent compound. The magnetic susceptibility measured along the three principal crystallographic directions for $Sm_{0.6}Y_{0.4}MnO_3$ and $Sm_{0.5}Y_{0.5}MnO_3$ deviate from a Curie-Weiss behavior below 100 K, which is more than 50 K above the Neel temperature $T_{N1} \sim 50$ K. Evidence for strong magnetoelectric coupling comes from the observation of coincident features in the magnetic susceptibility and dielectric constant data at $T_{N2} \sim 24$ K, and from the electric polarization which develops below this temperature. Equation 1 suggests that an electric polarization along the $c$ axis would be the result of cycloidal magnetic order in the $bc$ plane, propagating along the $b$ axis.[3,5] This expectation is confirmed by our neutron diffraction data that showed that for the Y doped samples there are two different magnetic phases found below 50 K. A sinusoidal order of the $Mn^{3+}$ moments along the $b$ axis below $T_{N1}$, and a cycloidal order of the $Mn^{3+}$ moments with the spins rotating in the $bc$ plane below $T_{N2} = 27$ K (Figure 12). Further evidence for a strong magnetoelectric coupling in these compounds is provided by the polarization flop from the $c$ axis to the $a$ axis under the application of a magnetic field parallel to the $c$ axis. According to the discussion by Mostovoy,[4] a polarization flop can be caused by the magnetic spiral changing orientation under a sufficiently strong magnetic field, with the spins eventually rotating about the magnetic field direction. For the case of $H \parallel c$, the spins would rotate in the $ab$ plane. Again referring to Equation 1, the polarization direction in this case would be parallel to the $a$ axis (i.e. $b \times (a \times b)$), as we have observed. A similar magnetic field dependence of the electric polarization was also found for $Eu_{0.6}Y_{0.4}MnO_3$, where a magnetic field applied parallel to the $a$ axis resulted in a polarization flop from the $a$ axis to the $c$ axis.[25] For both $Sm_{1-x}Y_xMnO_3$ ($x = 0.4, 0.5$) and $Eu_{0.6}Y_{0.4}MnO_3$, the polarization observed along the flop direction is considerably smaller than that seen in $TbMnO_3$.

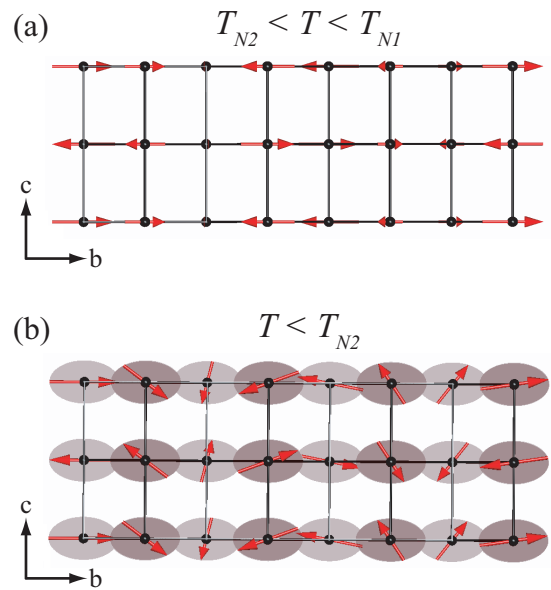

FIG. 12: The magnetic structure proposed for the $Mn^{3+}$ moments of $Sm_{0.6}Y_{0.4}MnO_3$ in the temperature ranges (a) $T_{N2} < T < T_{N1}$ and (b) $T < T_{N2}$.

Neutron diffraction data for $SmMnO_3$ confirm that the magnetic structure is that of an $A$-type antiferromagnet. Studies of orthorhombic $YMnO_3$ have indicated an incommensurate sine-modulated structure down to the lowest temperatures,[26] with a propagation vector $\mathbf{k} \sim (0\ 0.43\ 0)$ similar to that of $HoMnO_3$.[27] This structure does not transform into the $E$-type phase, as it does not show any locking of the propagation vector to the commensurate value $\mathbf{k} = (0\ \frac{1}{2}\ 0)$. We suggest that similar neutron diffraction studies of other compounds of the $RMnO_3$ series are required to clarify the phenomenology that was derived, for instance, from recent studies of the macroscopic properties of the Y-doped $EuMnO_3$ system. Such studies should provide a better understanding of the phase diagrams of the $RMnO_3$ manganites, in which rather different microscopic mechanisms are invoked to understand the multiferroic behavior in compounds with cycloidal[4] and $E$-type magnetic ground states.[9]

Our study of $Sm_{1-x}Y_xMnO_3$ shows that this system is similar in some respects to $TbMnO_3$. In both cases, the magnetoelectric coupling occurs below the magnetic transition temperature $T_{N2}$, and the characteristics of the magnetic order in the ferroelectric state (propagation vector, magnetic moments) are the same. In contrast to $TbMnO_3$, however, there is no evidence in our data to suggest that the rare earth magnetism in $Sm_{1-x}Y_xMnO_3$ plays a role in influencing the magnetic ordering of the $Mn^{3+}$ moments. The behavior of the dielectric constant and electric polarization of $Sm_{1-x}Y_xMnO_3$ in an applied magnetic field also contrasts with that seen in $TbMnO_3$. In this respect, our results on $Sm_{1-x}Y_xMnO_3$ should be compared with those obtained in multiferroic Y-doped $EuMnO_3$,[11,28,29] where similar changes in the direction of the polarization and the $Mn^{3+}$ moments in an applied magnetic field are observed.

# ACKNOWLEDGMENTS

D.O. would like to thank D.W. Baker for invaluable assistance with the refinement of the XRD data. We would like to acknowledge the help received from W. Kockelmann during experiments on GEM, ISIS (UK). D.O. and G.B. thank the STFC (UK) for the purchase of the isotopically enriched $Sm_2O_3$ powder for this experiment. This project was supported by funding from the EPSRC, UK. Some of the equipment used for these measurements was obtained through the Science City Advanced Materials: Creating and Characterising Next Generation Advanced Materials Project, with support from Advantage West Midlands (AWM) and part funded by the European Regional Development Fund (ERDF).